\global\def\draftcontrol{0}

   \def\versionno{ Framing }

\catcode`\@=11

\expandafter\ifx\csname draftcontrol\endcsname\relax\global\def\draftcontrol{0} 
\fi 

{\count255=\time\divide\count255 by 60 
\xdef\hourmin{\number\count255} 
\multiply\count255 by-60\advance\count255 by\time 
\xdef\hourmin{\hourmin:\ifnum\count255<10 0\fi\the\count255}} 
\def\draftdate{\number\month/\number\day/\number\year\ \ \ \hourmin } 

\newcommand\makepapertitle{\par

  \begingroup 
    \renewcommand\thefootnote{\@fnsymbol\c@footnote}%
    \def\@makefnmark{\rlap{\@textsuperscript{\normalfont\@thefnmark}}}%
    \long\def\@makefntext##1{\parindent 1em\noindent 
            \hb@xt@1.8em{%
                \hss\@textsuperscript{\normalfont\@thefnmark}}##1}%
     \newpage 
     \global\@topnum\z@   
     \@makepapertitle 
     \thispagestyle{empty}\@thanks 
  \endgroup 
  \setcounter{footnote}{0}%
  \global\let\thanks\relax 
  \global\let\makepapertitle\relax 
  \global\let\@makepapertitle\relax 
  \global\let\@thanks\@empty 
  \global\let\@author\@empty 
  \global\let\@date\@empty 
  \global\let\@title\@empty 
  \global\let\title\relax 
  \global\let\author\relax 
  \global\let\date\relax 
  \global\let\and\relax 
  \def\version{\let\version\@version\@gobble} 
} 
\def\@makepapertitle{%
  \newpage 
   \ifnum\draftcontrol=1 {} 
   \version\versionno 
   \vskip 5.5em%
   \else 
   \hfill\hbox to 3.5cm {\parbox{5cm}{\@pubnum}\hss}%
   \vskip 6.5em%
   \fi 
   \begin{center}%
   \let \footnote \thanks 
      {\hskip -0\textwidth \hbox to 1\textwidth%
        {\centerline{\Large\bf{\noindent%
	\parbox[t]{1.3\textwidth}{\begin{center}\@title\end{center}}}}}}%
     \vskip 1.5em%
     {\normalsize
       \lineskip .5em%
       \begin{tabular}[t]{c}%
         \@author 
       \end{tabular}\par}%
     \vskip 1.5em%
     {\@bstract}%
     \end{center}%
     \vfill
     \@date%
     \vskip 1.5em%
   \par 
} 

\gdef\@pubnum{} 
\def\pubnum#1{%
  \gdef\@pubnum{#1}} 

\gdef\@bstract{} 
\def\Abstract#1{%
  \gdef\@bstract{%
   \parbox{\textwidth-0pc}{%
   \centerline{\bf Abstract}\penalty1000 
   \noindent
   \renewcommand\baselinestretch{1.0} 
   {#1}}} 
} 

\gdef\@email{}
\def\email#1{%
   \gdef\@email{%
   Email: {\tt #1}}
}

\def\ps@paper{\let\@mkboth\@gobbletwo%
     \ifnum\draftcontrol=1 
        \def\@oddfoot{\hbox to \textwidth{\tiny \versionno \hfil\tiny\draftdate}%
        \hskip -\textwidth \hbox to \textwidth{\hfil\rm\thepage\hfil}}%
     \else\def\@oddfoot{\hbox to \textwidth{\hfil\rm\thepage\hfil}} 
     \fi 
     \let\@evenfoot\@oddfoot 
} 

\def\body{\clearpage 
          \pagestyle{paper} 
        } 
\newenvironment{acknowledgments}{%
\vskip 3.25ex 
\addcontentsline{toc}{section}{Acknowledgments}
\noindent {\bf Acknowledgments} 
} 

\def\@version#1{\ifnum\draftcontrol=1 
\typeout{}\typeout{#1}\typeout{} 
\vskip3mm\centerline{\hbox{\fbox{\normalsize{\tt DRAFT -- #1 -- } 
                   {\draftdate}}}}\vskip3mm 
\fi} 
\let\version\@version 
\long\def\eqlabel#1{\ifnum\draftcontrol=1 
                    \tag@false  
                    \tag*{(\theequation) \hbox to -0.2cm{\hspace{0cm}\small{#1}\hss}} 
                    \refstepcounter{equation}  
                    \edef\@currentlabel{\theequation} 
                    \ltx@label{#1}          
                    \else 
                    \label{#1} 
                    \fi 
                    } 
\let\st@bibitem\@bibitem 
\let\st@lbibitem\@lbibitem 
\ifnum\draftcontrol=1 
  \def\@bibitem#1{%
    \st@bibitem{#1}\a@@label{#1}\ignorespaces}
  \def\@lbibitem[#1]#2{%
    \st@lbibitem[#1]{#2}\a@@label{#2}\ignorespaces} 
  \def\a@@label#1{%
    \gdef\a@lab{\smash{\normalfont\small#1}} 
    \ifvmode 
      \if@inlabel 
        \global\setbox\@labels\hbox{%
          \llap{\a@lab\let\a@lab\relax 
                \kern\@totalleftmargin\kern\marginparsep}%
          \box\@labels}%
      \fi 
    \fi} 
\fi 

\documentclass[12pt,letterpaper]{article} 

\usepackage{amsmath,bm,amsfonts,amssymb,array,calc,amsthm,rotating}
\usepackage{epsfig,psfrag} 
\usepackage{amsthm}
\usepackage{graphicx}
\usepackage{color}
\usepackage[colorlinks=true]{hyperref}

\renewcommand\baselinestretch{1.25} 
\renewcommand\section{\@startsection {section}{1}{\z@}%
                                   {-3.5ex \@plus -1ex \@minus -.2ex}%
                                   {2.3ex \@plus.2ex}%
                                   {\normalfont\large\bfseries}} 
\renewcommand\subsection{\@startsection{subsection}{2}{\z@}%
                                   {-3.25ex\@plus -1ex \@minus -.2ex}%
                                   {1.5ex \@plus .2ex}%
                                   {\normalfont\normalsize\bfseries}} 
\renewcommand\subsubsection{\@startsection{subsubsection}{3}{\z@}%
                                   {-3.25ex\@plus -1ex \@minus -.2ex}%
                                   {1.5ex \@plus .2ex}%
                                   {\normalfont\normalsize\it}} 
\renewcommand\paragraph{\@startsection{paragraph}{4}{\z@}%
                                   {-3.25ex\@plus -1ex \@minus -.2ex}%
                                   {1.5ex \@plus .2ex}%
                                   {\normalfont\normalsize\bf}} 
\renewcommand\subparagraph{\@startsection{subparagraph}{5}{\z@}%
                                   {-1.25ex\@plus -1ex \@minus -.2ex}%
                                   {0ex \@plus .2ex}%
                                   {\normalfont\normalsize\it}}

\newtheorem{theorem}{Theorem}

\newenvironment{remark}{{\bf Remark}:}{\vskip 5mm }

\numberwithin{equation}{section}

\long\def\@makecaption#1#2{%
  \vskip\abovecaptionskip
  \sbox\@tempboxa{{\bf #1:} #2}%
  \ifdim \wd\@tempboxa >\hsize
    {\small\bf #1:} {\small #2}\par
  \else
    \global \@minipagefalse
    \hb@xt@\hsize{\hfil\box\@tempboxa\hfil}%
  \fi
  \vskip\belowcaptionskip}

\renewcommand*\l@section[2]{%
  \ifnum \c@tocdepth >\z@
    \addpenalty\@secpenalty
    \addvspace{.5em \@plus\p@}%
    \setlength\@tempdima{1.5em}%
    \begingroup
      \parindent \z@ \rightskip \@pnumwidth
      \parfillskip -\@pnumwidth
      \leavevmode \bfseries
      \advance\leftskip\@tempdima
      \hskip -\leftskip
      #1\nobreak\hfil \nobreak\hb@xt@\@pnumwidth{\hss #2}\par
    \endgroup
  \fi}
\renewcommand*\l@subsection{\addvspace{.0em \@plus\p@}\@dottedtocline{2}{1.5em}{2.3em}}
\renewcommand*\l@subsubsection{\addvspace{-.2em \@plus\p@}\@dottedtocline{3}{3.8em}{3.2em}}

\def\hepth#1{\href{http://xxx.arxiv.org/abs/hep-th/#1}{{arXiv:hep-th/#1}}}

\def\math#1{\href{http://xxx.arxiv.org/abs/math/#1}{{arXiv:math/#1}}}

\def\arxiv#1#2{\href{http://xxx.arxiv.org/abs/#1}{{arXiv:#1 [#2]}}}

\definecolor{refcol}{rgb}{0.0,0.0,0.2}
\definecolor{eqcol}{rgb}{.2,0,0}
\definecolor{purple}{cmyk}{0,1,0,0}

\gdef\@citecolor{refcol}
\gdef\@linkcolor{eqcol}
\gdef\@urlcolor{refcol}
\def\colorlinkspurple{\gdef\@urlcolor{purple}}
\def\colorlinksblue{\gdef\@urlcolor{blue}}
\def\colorlinksred{\gdef\@urlcolor{red}}

\def\ie{{\it i.e.}}

\def\revise#1       {\raisebox{-0em}{\rule{3pt}{1em}}%
                     \marginpar{\raisebox{.5em}{\vrule width3pt\ 
                     \vrule width0pt height 0pt depth0.5em 
                     \hbox to 0cm{\hspace{0cm}{%
                     \parbox[t]{4em}{\raggedright\footnotesize{#1}}}\hss}}}}

\def\calc         {{\cal C}}

\def\calm         {{\cal M}} 
\def\caln         {{\cal N}}

\def\C      {{\mathbb C}}

\def\Q    {{\mathbb Q}} 
 
\def\Z          {{\mathbb Z}}

\def\del          {\partial} 
 
\def\ee           {{\it e}}

\def\tr           {{\rm Tr}}

\def\Li {{\mathop{{\rm Li}_2}}}
\def\li#1{{\mathop{{\rm Li}_{#1}}}}

\newcommand\topa[2]{\genfrac{}{}{0pt}{2}{\scriptstyle #1}{\scriptstyle #2}}

\def\sqr#1#2{{\vcenter{\vbox{\hrule height.#2pt   
 \hbox{\vrule width.#2pt height#1pt \kern#1pt 
 \vrule width.#2pt}\hrule height.#2pt}}}}

\catcode`\@=12 

\begin{document} 


\title{Framing the Di-logarithm (over $\Z$)%
\footnote{Contribution to Proceedings of String-Math 2012, Bonn}}

\pubnum{%
}
\date{June 2013}

\author{
Albert Schwarz$^{\dag}$, Vadim Vologodsky$^{\ddag}$, and Johannes Walcher$^{\#}$ \\[0.2cm]
\it $^{\dag}$Department of Mathematics,\\ 
\it University of California, Davis, California, USA \\[.1cm]
\it $^{\ddag}$Department of Mathematics,\\ 
\it University of Oregon, Eugene, Oregon, USA \\[.1cm]
\it $^{\#}$Departments of Physics, and Mathematics and Statistics,\\ 
\it McGill University, Montreal, Quebec, Canada}

\Abstract{
Motivated by their role for integrality and integrability in topological string theory, 
we introduce the general mathematical notion of ``$s$-functions'' as integral linear 
combinations of 
poly-logarithms. $2$-functions arise as disk amplitudes in Calabi-Yau D-brane backgrounds 
and form the simplest and most important special class. We describe $s$-functions in terms 
of the action of the Frobenius endomorphism on formal power series and use this description 
to characterize 2-functions in terms of algebraic K-theory of the completed power
series ring. This characterization leads to a general proof of integrality of the 
framing transformation, via a certain orthogonality relation in K-theory. We comment on 
a variety of possible applications. We here consider only power series with rational
coefficients; the general situation when the coefficients belong to an arbitrary algebraic 
number field is treated in a companion paper.}

\makepapertitle

\body

\version\versionno

\vskip 1em


\hfill\hbox to 6.8cm{\parbox[t]{6.8cm}{
\footnotesize{\it \ldots as if this function alone among all others 
possessed a sense of humor}
(D.~Zagier \cite{zagier})}\hss}

\section{Introduction}

A remarkable aspect of exact calculations in supersymmetric and 
topological quantum theories is the blending of discrete and analytic
information. What we mean is that while, on the one hand, the microscopic 
Lagrangian formulation of some given supersymmetric observable makes manifest 
the holomorphic dependence on the parameters, and can, in some cases, be
used to derive the behavior under certain duality transformations (or
under analytic continuation), the typical answers localize to finite sums (or 
at the most, finite-dimensional integrals) over configurations of classical 
supersymmetric solutions (BPS states). 

In other words, the expansion coefficients of the supersymmetric amplitude around 
the appropriate limit often admit an {\it a priori} perhaps unexpected interpretation 
as counting dimensions of certain vector spaces of BPS states (or, more generally, an 
index of some operator on such spaces). That, conversely, the generating functions 
for these dimensions have interesting analytic and modular properties, is a remarkable 
fact that can be understood, at least in part, as a consequence of the underlying 
duality symmetries of the microscopic formulation.

In these respects, supersymmetric partition functions are reminiscent of 
functions of interest in analytic number theory, and in fact there are many cases 
in which the two are very closely related. This has led to a number of results on, 
most notably, modular forms finding applications in diverse areas of mathematical 
physics related to supersymmetric field and string theories. This interest has also 
led to a number of new mathematical results. The ways in which the functions of 
interest are related to some geometric situation in both physics and number theory
are also often similar. The rampant speculation about the deeper meaning of such
coincidences is best restrained by pointing out that there remain large classes of 
very deep number theoretic functions (such as L-functions, $\zeta$-functions) 
whose relevance for supersymmetric quantum theory are much less clear.

In this paper, we study elementary algebraic properties of a certain class of functions 
that we call ``$s$-functions'' (where, at least for now, $s$ is a positive integer). We
extract this notion from the appearance of $s$-function in perturbative computations in 
topological string theory, where they are building blocks of supersymmetric generating 
functions. 
We define an $s$-function as an integral linear combination of 
$s$-logarithms. We give later an equivalent definition in terms of Frobenius map
on (formal) power series with rational coefficients. That second definition can easily
be generalized to the case  when we allow coefficients to lie in arbitrary local or 
global number fields, see \cite{svw2}.

One of our main results concerns the most important special case, $s=2$, and is
an integrality statement of a certain algebraic transformation of $2$-functions 
(viewed as formal power series) 
that we call ``framing''. We point out that while special cases of this
framing transformation are known in the context of open topological string theory 
(where we borrowed the name, see \cite{akv}), the generality in which it applies has not been 
pointed out in the literature to our knowledge (Although it might be known to experts.
We made our initial observations after reading \cite{stienstra}.) Secondly, we will 
give a mathematical proof of this framing property, using an interpretation  of the 
notion of $2$-function in algebraic K-theory. (M. Kontsevich informed us that he also 
obtained this interpretation and used it to prove some integrality theorems.)

Finally, we point out a few further generalizations of our setup and constructions.
One of these generalizations involves extending the field of definition of
the coefficients from $\Q$ to a more general number field. The relevance of 
such extensions was first observed in \cite{arithmetic}, and we will elaborate on 
them in a companion paper \cite{svw2}.

At the moment the only immediate applications of our results that we are aware of
come from open topological string theory and mirror symmetry. We suspect however 
that the concepts we introduce might play a role in other contexts as well. As a 
particular example, refs.\ \cite{frv,stienstra} lead us to expect certain connections 
with the theory of Mahler measures. In a different way, the characterization of 
2-functions in algebraic K-theory is reminiscent of a certain integrability
condition recently explored by Gukov and Sulkowski \cite{gusu}.

In mathematical terms, integrality of framing is the following statement. Say
\begin{equation}
\eqlabel{say}
W(z) = \sum_{d=1}^\infty n_d\Li(z^d)
\end{equation}
is an integral linear combination of standard di-logarithms,
\begin{equation}
\eqlabel{dilog}
\Li (z) = \sum_{k=1}^\infty \frac{z^k}{k^2}
\end{equation}
Namely, the coefficients $n_d$ in \eqref{say} are integers, and we may as well assume 
that the series \eqref{say} is convergent. We introduce the power series
\begin{equation}
\eqlabel{introduce}
Y(z) = \exp\bigr(-z \frac{d}{dz}W(z)\bigr) \,,
\end{equation}
with constant term $1$ around $z=0$. Then the relation
\begin{equation}
\eqlabel{how1}
\tilde z = -z Y(z)
\end{equation}
may be inverted (formally, and as a convergent power series),
\begin{equation}
\eqlabel{how2}
z = - \tilde z \tilde Y(\tilde z)
\end{equation} 
to yield another power series $\tilde Y(\tilde z)$ around $\tilde z=0$ with 
constant term $1$. Expanding
\begin{equation}
\tilde W(\tilde z) = 
\int\frac{d\tilde z}{\tilde z} \bigl(-\log\tilde Y(\tilde z)\bigr)
= \sum_{d=1}^\infty \tilde n_d \Li (\tilde z^d)
\end{equation}
defines coefficients $\tilde n_d$ which are {\it a priori} rational numbers.
It is elementary to see that $\tilde Y(\tilde z)$ has integer coefficients,
and as a consequence that $d\,\tilde n_d\in \Z$. We claim that, in fact,
\begin{equation}
\eqlabel{muchmore}
\tilde n_d\in \Z
\end{equation}
(whenever $n_d\in\Z$). See section \ref{proof} for the sketch of a proof of this statement.

In physical terms, we think of $W(z)$ as a contribution to the space-time 
superpotential from D-branes wrapping supersymmetric cycles in some Calabi-Yau
compactification of string theory preserving $\caln=1$ supersymmetry in 4
dimensions. In that context, $z$ is a chiral superfield whose vacuum expectation
value parametrizes
the moduli space of open/closed string vacua, and corresponds to a geometric
modulus of the configuration. It was shown long time ago by Ooguri and Vafa 
\cite{oova}, generalizing work by Gopakumar and Vafa \cite{gova}, that, for an
appropriate choice of parametrization, one 
expects the coefficients $n_d$ in the expansion \eqref{say} to count dimensions 
of spaces of appropriate BPS states, hence the integrality.

One of the features of the setup of Ooguri and Vafa is the dependence of
the superpotential (and the BPS invariants) on an integer parameter, $f$, 
known as ``the framing''. Algebraically, the framing results from an ambiguitiy 
in the identification of the open string modulus \cite{akv}. Namely, framing
by $f$ amounts to replacing
\begin{equation}
\eqlabel{framing}
\log z \to \log z_f = \log z + f z\del_z W(z)  \,,\qquad f\in\Z
\end{equation}
Although it might not be immediately obvious, the transformations \eqref{framing} 
are very closely related to (namely, generated by) \eqref{how1}, \eqref{how2}.
We will explain this connection, and also review the general setup in 
somewhat more detail, in section \ref{OV}.

The dependence on $f$ of open topological string 
amplitudes in the Ooguri-Vafa setup is explained through the (large-$N$) 
duality with Chern-Simons theory and knot invariants. (The framing of a knot
is the choice of non-vanishing section of the normal bundle of a knot in
a three-manifold, and gets identified with $f$ under this duality.) 
For this duality to 
operate, it is important that the underlying Calabi-Yau manifold be {\it non-compact}. 
Mathematically, the framing dependence of the enumerative (open Gromov-Witten)
invariants is only well understood when the manifold is {\it toric}
\cite{kali}.

In a series of works, from \cite{opening} to \cite{arithmetic}, it was shown 
that the expansion \eqref{say} continues to be valid in principle when the
underlying Calabi-Yau manifold is {\it compact}, and {\it non-toric}, but 
requires a number of modifications in practice. Most prominently, the parameter 
$z$ should be a closed string modulus that remains massless at tree-level.
Secondly, the standard di-logarithm \eqref{dilog} has to be ``twisted'' in general to 
take into account the symmetries of the open string vacuum structure. 
On the other hand, neither the geometric setup (in the A-model) nor the actual 
calculation (in the B-model) seem to involve any ambiguity that could be identified 
as ``framing''. 
 Most recently, it was pointed out in \cite{arithmetic} that the 
generic B-model setup will predict invariants $n_d$ that are {\it irrational} 
numbers valued in some algebraic number field (a finite extension of $\Q$,
fixed for each geometric situation). The symmetry of the space of vacua 
and the associated twist of the di-logarithm was related to the Galois group of 
that finite field extension.  It is possible to prove some integrality statements 
also in this case (see \cite {svw2} for detail).

The statement that D-brane superpotentials given by geometric formulas of
\cite{agva,akv,opening} indeed admit a decomposition of the form \eqref{say}
into integral pieces was proven mathematically in ref.\ \cite{sv2}, using
and extending earlier work by the same authors \cite{ksv,sv1}. The central
aspect of that series of works was to relate the BPS numbers $n_d$ to the
action of the Frobenius automorphism on $p$-adic cohomology of the Calabi-Yau
(together with an algebraic cycle).

The proofs of \cite{sv2} show integrality of open topological string amplitudes,
separately for any value of the framing, in situations in which this concept is
well-defined. One of the main messages of the present paper is that the integrality of 
the framing transformation is more general, and in fact not tied to a particular 
geometric siutation. But the methods for proving the integrality statements developed
in \cite{sv2} continue to apply. This means in particular that we can define a ``framed''
superpotential and enumerative invariants even when we do not know a geometric 
interpretation for the integer ambiguity $f$. We interpret this fact, together 
with the observation that instanton numbers are related to Mahler measures by a 
framing transformation \cite{frv,stienstra}, as a hint that framing is an important
intrinsic property of the di-logarithm.

As further support, we mention that the integrality of the framing transformation 
is naturally expressed as a certain torsion/orthogonality condition in algebraic 
K-theory, see section \ref{proof}, and can also be given a Hodge theoretic 
interpretation. Framing can also be generalized to the multi-variable situation, 
in which it depends in an interesting way on the additional data of a symmetric
bilinear form.
Finally, while in this paper we are concerned mainly with the situation in which 
the coefficients are actually rational numbers, the generalization to arbitrary number 
fields is rather straightforward. We will consider it in a separate paper \cite{svw2},
where further mathematical details may also be found.

\section{Dilogarithm, \texorpdfstring{$s$}{s}-functions, and topological strings}
\label{OV}

(The mathematically inclined reader may gain from skipping the odd (numbered)
subsections
which provide some physics motivation for our definitions.)

\subsection{A-model}

From the point of view of the A-model, the origin of the formula \eqref{say} is,
intuitively, easy to understand. Consider a Calabi-Yau threefold $X$ and a 
Lagrangian submanifold $L\subset X$. We view $L$ as the support of a topological 
D-brane in the A-model, which we may want to use as an ingredient in a superstring 
theory construction. As is well-known, the classical deformation space of $L$ modulo 
Hamiltonian isotopy (or, preserving the ``special Lagrangian'' condition, if one exists) 
is unobstructed and of dimension equal to $b_1(L)$. Worldsheet instanton corrections
however induce a space-time superpotential that schematically takes the form
\begin{equation}
\eqlabel{WW}
W = \ldots + \sum_{u:(D,\del D)\to (X,L)} \ee^{\int_D u^* \omega}\,
\tr \bigl[ P\ee^{\oint_{\del D} u^* A}\bigr]
\end{equation}
and depends (via the symplectic form $\omega$) on the K\"ahler moduli of $X$ 
and (via the (unitary) connection $A$) on the choice of a flat bundle over $L$.
Here, $\ldots$ denotes certain (subtle) classical terms that we will 
neglect in this paper, so the sum is over all (non-constant) holomorphic maps
\begin{equation}
\eqlabel{maps}
u:(D,\del D)\to (X,L) 
\end{equation}
from the disk $D$ to $X$ mapping the boundary $\del D$ to $L$.

It is well known that the expected (virtual) dimension of the space of such holomorphic
maps is zero for any class $\beta = u_*([D,\del D]) \in H_2(X,L)$, and hence 
one expects to write\footnote{Note that the ${\rm Tr}$ in \eqref{WW}
really depends on the homotopy class of $u$ in $\pi_2(X,L)$. The formula
\eqref{this} makes sense if the fundamental group of $L$ is abelian,
so that $\pi_2(X,L)=H_2(X,L)$.}
\begin{equation}
\eqlabel{this}
W(Q) = \sum_{\beta\in H_2(X,L)} m_\beta q^{\beta}
\end{equation} 
where $\log q$ is the appropriate combination of moduli of $(X,L)$, and $m_\beta$ is 
the ``number'' of holomorphic maps in a fixed class $\beta$ (open Gromov-Witten 
invariants).
While the general definition of $m_\beta$ is plagued with difficulty, 
it is in any case 
clear that the moduli space $\calm(\beta)$ of such maps will contain components
of positive dimension, if $\beta$ is not primitive. Namely, say $\beta=k \beta'$ 
with $\beta'$ integral and $k>1$. Then any $u'\in \calm(\beta')$ may be
composed with a degree $k$ covering map $c:(D,\del D)\to (D,\del D)$ to give
a map $u=u'\circ c\in\calm(\beta)$. Since the maps $c$ come in families (of
dimension $2k-2$), so $\calm(\beta)$ will contain components of positive
dimension.

The formula \eqref{say} is a reflection of these multi-covers. (Even though it
might not be strictly true that all holomorphic maps can be factorized in this
way, see \cite{lazzarini}, the success of the formula suggests that this is effectively the case.)
The general statement is that a BPS state corresponding (intuitively) to an 
embedded disk with boundary on $L$ in the class $\beta$, together with all its 
multi-covers, makes a contribution to $W$ of the form
\begin{equation}
\eqlabel{proto}
W_\beta(q) = \sum \frac{1}{k^2} q^{k\beta} = \Li(q^\beta) \,,
\end{equation}
so that if $n_\beta$ is the (integer!) degeneracy of BPS states of charge
$\beta$, the total superpotential is
\begin{equation}
\eqlabel{total}
W(q) = \sum_\beta n_\beta W_\beta(q) = \sum_\beta n_\beta \Li(q^\beta)
\end{equation}
Eq.\ \eqref{say} is recovered when $H_2(X,L)$ has rank one, and $q=z$.

The prototypical example of a multi-cover formula like \eqref{proto} is
known from the beginning of mirror symmetry \cite{cdgp} as the
Aspinwall-Morrison formula \cite{asmo}. It states that the large volume
(A-model) expansion of the $\caln=2$ prepotential (i.e., the genus $0$ 
Gromov-Witten potential) takes the form
\begin{equation}
\eqlabel{type}
F^{(0)} = \sum M_d q^d = \sum N_d \li{3}(q^d)
\end{equation}
with {\it integer} $N_d$. The $M_d$ are rational numbers, which 
is obvious from both the definition in Gromov-Witten theory, as well
as from the B-model formulas (involving differential equations with
rational coefficients). The integrality of the $N_d$ however is harder to
see. It was proven mathematically in \cite{ksv,sv1,sv2,vadim}. A physical
explanation was given in \cite{gova} by relating the $N_d$ to the degeneracy
of BPS states. The generalization of the Aspinwall-Morrison formula to 
arbitrary genus $g$ was shown to involve the poly-logarithm $\li{3-2g}$.

\subsection{\texorpdfstring{$s$}{s}-functions}

The central importance of these multi-cover formula in relating the perturbative
topological string amplitudes to the degeneracy of BPS states motivates us
to introduce the following notion: If $s$ is a positive integer, we call a power 
series
\begin{equation}
V (z)= \sum_{d=1}^\infty m_d z^d  \in \Q[[z]]
\end{equation}
with rational coefficients $m_d$ an $s$-function if it can be written as 
an integral linear combination of $s$-logarithms.
\begin{equation}
V(z) = \sum_{d=1}^\infty n_d \li{s}(z^d)\,,\qquad \li{s}(z) := \sum_{k=1}^\infty
 k^{-s} z^k
\end{equation}
with $n_d\in\Z$. It is convenient to define the logarithmic derivative,
\begin{equation}
\delta_z = \frac{d}{d\ln z}
\end{equation}
So $\delta_z \li{s}(z) = \li{s-1}(z)$, and if $V(z)$ is an $s$-function, 
$\delta_z V(z)$ is an $(s-1)$-function.
For the topological string, the relevant values are $s=3$ for genus $0$ 
(closed string tree-level) invariants, $s=2$ for disk invariants (open string
tree-level), and $s=1$ for all one-loop amplitudes (open or closed).

Sometimes it is convenient to consider $s$-functions with respect to prime number 
$p$ requiring that the denominators of  the coefficients $m_d$ are not divisible 
by $p$ (the coefficients are $p$-integral). 

\subsection{B-model}

As mentioned in the introduction, framing originally entered topological string 
theory through the relation between local toric manifolds and Chern-Simons gauge
theory and knot invariants. Framing has also been explained in (toric) A-model \cite{kali} 
as a choice of linearization of the torus action required to make the localization 
calculation of open Gromov-Witten invariants well-defined. The operation itself is 
however most straightforward to explain geometrically in the B-model.

The B-model mirror of a general toric Calabi-Yau threefold has the form
\begin{equation}
\eqlabel{Bmodel}
\{ u v = H(x,y) \} \subset \C\times\C\times\C^*\times\C^*\ni (u,v,x,y)
\end{equation}
where we do not need to write explicitly the dependence on complex structure
parameters. Geometrically, \eqref{Bmodel} is a conic bundle over $\C^*\times\C^*$
with discriminant locus given by the (non-compact, \ie, punctured) curve
\begin{equation}
\calc = \{ H(x,y)=0\} \subset \C^*\times\C^*
\end{equation}
Aganagic and Vafa \cite{agva} study B-type D-branes in this geometry wrapped on 
one component of a reducible fiber, say $u=0$, varying over $\calc$. They identify 
certain ``semi-classical'' regimes of these branes as punctures of the curve, and 
show that the superpotential expanded near such a point is given by an
Abel-Jacobi computation on $\calc$. For simplicity, let us say the interest is
in a puncture at $y=1$. Then the superpotential is given by the formula
\begin{equation}
\eqlabel{formula}
\delta_z W(z) = - \log y(z) 
\end{equation}
where $z$ (the open string modulus) is a local coordinate on the curve such that $z=0$ 
corresponds to the puncture. This choice is made such that the superpotential
is critical, \ie, $y=1$, at $z=0$. In some simple cases, $z$ coincides with $x$
in \eqref{Bmodel}.)
As pointed out in \cite{akv}, however, this 
prescription is ambiguous: If $z$ is such a good coordinate, then so is any combination
\begin{equation}
\eqlabel{ambi}
z_f = z (-1)^f y^f
\end{equation}
with integer $f$. Up to the sign, this is equation \eqref{framing} from the 
introduction.\footnote{The sign is thrown in to preserve integrality also
at $p=2$, see below.}

Let us pause briefly here to explain the relevance of the mirror map: $W$ as defined 
by \eqref{formula} is a 2-function in the sense of the previous subsection when 
expanded, not only around $z=0$, but also in the appropriate flat closed string 
coordinates around a degeneration of the curve. Assuming that the family of curves 
is defined over $\Q$, the 2-function property of $W$ (after the mirror map) was 
shown in general in \cite{sv2}. However, as pointed out in general
in \cite{akv}, the mirror map is in fact independent of the open string coordinate 
itself, and as a consequence framing commutes with the mirror map. This observation
also explains why we are using the traditional B-model notation $z$ interchangeably 
with the A-model $q$ for the argument of our $s$-functions.

Our main point now is to abstract the ambiguity \eqref{ambi} to the following
general ``framing transformation'', parameterized by an integer $f$. 

\subsection{The group of framing transformations}

Say $V(z)$ is an $s$-function with $s\ge 1$. Define 
\begin{equation} 
\eqlabel{witt}
Y = \exp\bigl(-(\delta_z)^{s-1} V\bigr) \in \Z[[z]]
\end{equation}
(The integrality of the power series follows from $\delta_z^{s-1}\li{s}(z) = 
\li{1}(z) = -\ln(1-z)$.)
When $s=2$ and $V=W$ is the superpotential of (the mirror of) a toric D-brane configuration, 
then the corresponding $Y = \exp(-\delta_z W)=y(z)$ will by the above construction satisfy 
an algebraic equation
\begin{equation}
\eqlabel{algebraic}
H(z,Y) = 0
\end{equation}
while the ``framed superpotential'' $\delta_{z_f}W_f=-\log Y_f$ can be identified with a solution 
of the equation
\begin{equation}
H_f(z_f,Y_f) = H(z_f (-Y_f)^{-f}, Y_f) = 0
\end{equation}
Even though the equations are algebraic, it is natural (for the purposes of
mirror symmetry, for example) to think of solutions $Y(z)$ and 
$Y_f(z_f)$ as local power series around $z=0$, $z_f=0$, respectively. 
By the construction, both $Y(0)=1$, and $Y_f(0)=1$. The relation
between the two is then simply $Y_f(z_f)=Y(z)$, $z_f=(-Y)^f z$. Eliminating $z$,
and renaming $z_f$ as $z$, this means that we can obtain $Y_f(z)$ as the solution 
to the equation in formal power series,
\begin{equation}
\eqlabel{or}
Y_f = Y\bigl(z (-Y_f)^f\bigr)
\end{equation}
The ``framed 2-function'' is then the power series
\begin{equation}
\eqlabel{int}
W_f(z) = \int \frac{dz}{z} \bigl(-\log Y_f(z)\bigr)
\end{equation}
and, as we prove below, in fact is also a 2-function. Since the relations
\eqref{or}, \eqref{int} make sense independent of the existence of an algebraic
equation of the type \eqref{algebraic}, we may take this as the general definition
of framing, even for cases in which no such equation is known to exist.

We note a few elementary properties of this definition. First of all, framing defines 
a group action $\Z\ni f: W\to W_f$. Indeed, using the definitions, we find
\begin{equation}
\begin{split}
(Y_f)_{f'} &= Y_f\bigl(z \bigl(-(Y_f)_{f'}\bigr)^{f'}\bigr) \\
&= Y\bigl(z\bigl(-(Y_f)_{f'}\bigr)^{f'} \bigl(-(Y_f)_{f'}\bigr)^{f}\bigr) \\
&= Y\bigl(z \bigl(-(Y_f)_{f'}\bigr)^{f+f'}\bigr)
\end{split}
\end{equation}
Thus, the equation for $(Y_f)_{f'}$ is exactly the defining equation for $Y_{f+f'}$.

Secondly, to make contact with the introduction, we write eqs.\ 
\eqref{how1}, \eqref{how2} in the form
\begin{equation}
z= - \tilde z \tilde Y = z Y\bigl(-\tilde z\tilde Y\bigr) \tilde Y
\end{equation}
which is equivalent to
\begin{equation}
Y(-\tilde z\tilde Y) = \tilde Y^{-1}
\end{equation}
Comparison with \eqref{or} shows that
\begin{equation}
\tilde Y = (Y_{-1})^{-1}
\end{equation}
By substituting $Y_1$ for $Y$ into this equation, we learn that $\widetilde {Y_1}=
Y^{-1}$, and since $Y\mapsto \tilde Y$ is obviously involutive, this implies
\begin{equation}
Y_1 = \widetilde{Y^{-1}} 
\end{equation}
We conclude that framing transformations in the sense of \eqref{or}
in fact are generated by the single
transformation \eqref{how1}, \eqref{how2}, together with the operation $Y\mapsto
Y^{-1}$ (which in terms of the 2-function, corresponds simply to 
$W\mapsto -W$). Hence, we will refer to \eqref{how1}, \eqref{how2} also as ``framing''.

\section{\texorpdfstring{\mbox{K-theoretical description of 2-functions. Integrality of 
framing}}{K-theoretical description of 2-functions. Integrality of framing}}
\label{proof}

The purpose of this section is to  give a description of 2-functions in terms of 
K-theory and to derive from this description the integrality of the framing 
transformation:

\begin{theorem}
\label{intthm}
If $W(z)\in \Q[[z]]$ is a 2-function , then
its image under framing, $\tilde W(z)$ is also a 2-function . 
\end{theorem}
To prove this theorem it is sufficient to check for all prime numbers $p$ that the 
fact that $W(z)$ is a 2-function with respect to $p$ implies that  $\tilde W(z)$ 
is also a 2-function with respect to $p$.

A  complete proof of this statement  will be given in \cite {svw2}. The $K$-theoretic 
proof that we  will sketch  in this section works only for odd primes.

\noindent
We begin with a lightning review of algebraic K-theory.

\subsection{Algebraic K-theory. Orthogonality relation}

\newcommand{\GL}{{\it GL}}
For a ring $A$, the group $K_1(A)$ 
is defined as the abelianization of the infinite linear group $\GL(A)$: 
\begin{equation}
K_1(A)=\GL(A)/[\GL(A),\GL(A)].
\end{equation}
If $A$ is a Euclidean domain (in particular, it is a commutative ring), then the
group $K_1(A)$ is isomorphic to the group $A^{\times}$ of invertible elements of 
$A$. (For arbitrary commutative rings we have an embedding  $A^{\times}\to K_1(A)$ 
induced by the embedding $A^{\times}=\GL_1(A)\hookrightarrow \GL(A)$, and a map 
$K_1(A)\to A^{\times}$ 
induced by the determinant map $\det:\GL(A)\to A^{\times}.$)

Notice that $K_1(A)$ is usually regarded as an additive group, but in our situation
it is isomorphic to a multiplicative group $A^{\times}$, and so the multiplicative 
notation is more convenient.

The group $K_2(A)$, which we will write {\it additively}, is defined for an arbitrary ring
via the universal central extension of the commutator subgroup $E(A)=[\GL(A),\GL(A)]$. Thus
it fits into the sequence
\begin{equation}
K_2(A) \to {\rm St}(A) \to E(A) \to \GL(A) \to K_1(A)
\end{equation}
where ${\rm St}(A)$ is the Steinberg group (when $A$ is Euclidean, we may think of
the ``universal cover'' of $E(A)={\it SL}(A)={\rm Ker}(\det)$).

For an arbitrary ring $A$ there exists a pairing $K_1(A)\otimes_\Z K_1(A)\to K_2(A)$. 
Via the embedding $A^\times\to K_1(A)$, this pairing induces a (skew) pairing of 
invertible elements of $A$ in $K_2(A)$ (an antisymmetric bilinear map $\phi: 
A^{\times}\otimes A^{\times}\to K_2(A)$).

An important property of $\phi$ is that the pairing of two invertible elements 
$f,g$ vanishes if $f+g=1$.
Let us denote by $J$ the subgroup of $A^{\times}\otimes A^{\times}$ 
generated by elements of the form $f\otimes (1-f);$ the above statement 
means that $J\subset {\rm Ker}\phi.$ If $A$ is a field, then  $J= {\rm Ker}\phi$
(by Matsumoto's theorem).

Using the notation 

$$K^0_2(A)= A^\times\otimes A^\times/J$$

we can consider the pairing $\phi$  as a composition of maps $ A^\times\otimes 
A^\times\to K^0_2(A)$ and $K^0_2(A)\to K_2(A)$.  We will work with the first map 
considered as a pairing on $A^\times$; we denote this pairing by $\{f,g\}$. (It 
is possible to work also with $\phi$, but this makes the proof more complicated.)

We will characterize 2-functions in terms of this pairing.

Let us take two invertible elements $f\in A^{\times},g\in A^{\times}$. By definition $f$ and 
$g$ are orthogonal if the element $2\{f,g\}=\{f^2,g\}=\{f,g^2\}$ 
vanishes. We have used the bilinearity of the pairing $\phi$  expressed by formulas 
$\{f_1f_2,g\}=\{f_1,g\}+\{f_2,g\}, \{f,g_1g_2\}=\{f,g_1\}+\{f,g_2\}.$ These unusual 
formulas come from the fact that the operation in $A^{\times}$ is written as multiplication,
while the operations in  $K_2(A)$ and $K^0_2(A)$ are written additively. \footnote{Some 
other useful properties include
$\{f,-f\}=0$, $\{f,1\}=0$, and anti-symmetry $\{f,g\}=-\{g,f\}$.} 
Notice that it follows from bilinearity that an invertible element $f$ that is orthogonal 
to elements $g\in A^{\times}$ and $h\in A^{\times}$ is orthogonal to their product $gh.$  
It is also obvious 
that automorphisms of the ring $A$ preserve orthogonality.

The relevant ring for us is $A=\Z((q))$, the ring of formal Laurent series in one variable
$q$, with integer coefficients. The ring $A$ has a natural topology, that induces a 
topology in $A^{\times}.$ This allows 
us to modify the notion of orthogonality: we will say that $f,g$ are orthogonal in the new 
sense if there exists a sequence of pairs $(f_n,g_n)$ such that $f_n$ tends to $f$, $g_n$ 
tends to $g$ and $f_n$ is  orthogonal to  $g_n$ in the old sense . 

Notice  that starting with a 2-function 
$W(q)=\sum_{d=1} n_d \Li(q^d)$
represented as a  sum of di-logarithms we construct an invertible element of $A$ by the formula
$$Y(q)=\exp (-\delta_qW(q))=\prod (1-q^d)^{dn_d}.$$

If the sum of di-logarithms is finite the element $Y(q)$ is orthogonal to $q$. It  is 
sufficient to check this statement for every factor.  The fact that  $q$ is orthogonal 
to $(1-q^d)^d$ can be derived from the following chain of identities:
\begin{equation}
\label{ }
\{q,(1-q^d)^d\}=\{q^d,1-q^d\}=0.
\end{equation}  
If the sum of di-logarithms is infinite then $Y(q)$ is orthogonal to $q$ in the new 
sense (infinite product is a limit of finite products).  {\it In what follows we will 
understand the orthogonality in the new sense.} 

Let us notice that $q^m$ is orthogonal to $q.$ It is sufficient to check this for $m=1$.
This follows from the identiy
$$\{q, (-q)\}= \{q,(1-q)\} + \{q^{-1}, (1-q^{-1})\} =0,$$
which in turn follows from $ -q = \frac{1-q}{1-q^{-1}}.$
We see that  $2\{q, q\} = \{q,(-q)^2\} = 2\{q, (-q)\}= 0.$

We obtain that $q$ is orthogonal to an expression of the form $q^m\prod (1-q^d)^{dn_d}.$ 
One can prove  (Sec 5 and \cite {svw2} )  that the inverse statement is also correct:

\begin{theorem}
\label{equithm}
{\it If  $Y(q)=\exp (-\delta_qW(q))$
is orthogonal to $q$ then  the series $W(q)\in \Q[[q]]$ is a 2-function for all odd primes .}
\end{theorem}

In other words $q$ is orthogonal to $Y$  where $Y$ behaves like $q^m$ as $q$ tends to $0$
if and only if $\int \log( Y/q^m) d \log q$ is a 2-function for all odd primes.

\subsection{Integrality of framing (Proof of Theorem \ref{intthm})}

Notice that any change of variables $\tilde q=q+a_2 q^2+a_3 q^3+ \cdots$ with $a_i\in\Z$ induces
an automorphism of the algebra $A=\mathbb{Z}((q))$. This automorphism preserves the 
orthogonality relation. 

Using this fact we can describe all orthogonal pairs $(f,g).$ It is sufficient to consider 
only the case when $f$ behaves like $q$ as $q$ tends to zero. Then we can take $f$ as a 
new variable; in terms of this variable $g$ has an expression of the form 
$f^m\prod(1-f^d)^{dn_d}$ and integrates to a 2-function $\sum n_d \Li(f^d).$ 
This follows from Theorem \ref{equithm} above, applied to the ring $\Z((f))$,
which is isomorphic to $A$ as remarked above.

Let us now consider all orthogonal pairs $(f,g)$ where both $f$ and $g$ behave like $q$ as 
$q\to 0$. The orthogonality relation is symmetric, hence $f$ and $g$ are on equal footing. 
Therefore, we can construct two different 2-functions (one from expression of $g$ 
in terms of $f$, another from expression of $f$ in terms of $g$.)  These two 2-functions 
are related by framing transformation \eqref{how1}, \eqref{how2}. Thus we see that
Theorem \ref{intthm} is a simple consequence of the K-theoretical description of 2-functions
and of the symmetry of the orthogonality relation.

Notice that in the proof we used orthogonal pairs where both $f$ and $g$ behave like 
$q$ as $q\to 0 $ (this is important for symmetry). In the orthogonal pair $(q,Y(q))$ 
the function $Y(q)$ does not satisfy this condition, therefore in the construction of 
the framing transformation it should replaced by $qY(q).$

\section{Frobenius automorphism and local \texorpdfstring{$s$}{s}-functions}

The purpose of this section is to reformulate the definition of an $s$-function in 
terms of the action of the Frobenius endomorphism acting on (formal) power series.
Such a reformulation was crucial in the proofs of integrality theorems in  
\cite{ksv,sv1,sv2}. We will apply it here to sketch a proof of  the description of 2-functions 
in terms of algebraic K-theory (Theorem \ref{equithm}) that was used in the derivation
of integrality of framing (see \cite{svw2} for more detail).

First of all we will check that $V(z)\in \Q[[z]]$ is an $s$-function if and only
if for any prime number $p$ 
the formal series
\begin{equation}
\eqlabel{fr}
\frac{1}{p^s}(V(z^p)) - V(z) 
\end{equation}
is $p$-integral (the denominators of the coefficients are not divisible by $p$). 
The proof is an easy 
consequence of M\"obius inversion formula (and a trivial generalization of the special
statements for $s=2,3$).

Recall that if $(a_d)$ and $(b_d)$ are two sequences such that
\begin{equation}
\eqlabel{p1}
a_d = \sum_{k|d} b_k
\end{equation}
(where the sum is over all divisors of $d$), then
\begin{equation}
\eqlabel{p2}
b_d = \sum_{k|d} \mu({\textstyle\frac{d}{k}}) a_k = \sum_{k|d} \mu(k) a_{k/d}
\end{equation}
where $\mu$ is the M\"obius function: $\mu(k)=0$ if $k$ is not squarefree, $\mu(k)=(-1)^r$
if $k=p_1\cdots p_r$ is the product of $r$ distinct prime factors.
The important property of $\mu$ is that
\begin{equation}
\eqlabel{important}
\sum_{k|d} \mu(k) = \begin{cases} 1 &\text{if $d=1$}\\
0 &\text{if $d>1$}
\end{cases}
\end{equation}
which itself follows from the fact that if $r>0$
\begin{equation}
\sum_{l|k} \mu(l)=\sum_{l|p_2\cdots p_r} \mu(l) + \sum_{l|p_2\cdots p_r} \mu(p_1 l)
\end{equation}
Then \eqref{p2} follows from the computation
\begin{equation}
\sum_{k|d} b_k = \sum_{l|k|d} \mu({\textstyle\frac{k}{l}}) a_{l} = \sum_{l|d} a_l 
\sum_{\topa{k|d}{l|k}} \mu({\textstyle\frac{k}{l}}) = \sum_{l|d} a_l \sum_{k|\frac{d}{l}} \mu(k) 
= a_d
\end{equation}

Returning to $s$-functions, we compare coefficients of $z^d$ in
\begin{equation}
V(z) = \sum_{d=1}^\infty m_d z^d = \sum_{d,k=1}^\infty \frac{n_d}{k^s} z^{d k}
\end{equation}
to conclude that 
\begin{equation}
\eqlabel{that}
d^s m_d = \sum_{k|d} k^s n_{k}
\end{equation}
where the sum is over all divisors of $d$. Applying M\"obius inversion, we find
\begin{equation}
\eqlabel{thiss}
d^s n_d = \sum_{k|d} \mu({\textstyle\frac{d}{k}}) k^s m_k
\end{equation}
On the other hand, the statement that \eqref{fr} be $p$-integral for all primes $p$
is equivalent to the condition that
\begin{equation}
\eqlabel{condition}
m_{d}- \frac{1}{p^s} m_{d/p}  \quad\text{be $p$-integral for all $p$, $d$}
\end{equation}
it being understood that $m_{d/p}=0$ if $p\nmid d$.
To see that the integrality of the $n_d$ implies this condition, we note that 
\eqref{that} implies
\begin{equation}
m_{d} - \frac{1}{p^s} m_{d/p} = \frac{1}{d^s} \sum_{\topa{k|d}{k\nmid d/p}} k^s n_k
\end{equation}
The sum is restricted to those $k$ divisible by as many powers of $p$ as $d$, 
and therefore the right hand side is $p$-integral if the $n_k\in\Z$.

Conversally, since $\mu(k)=0$ if $k$ is divisible by $p^2$, and $\mu(pk)=-\mu(k)$ 
if $p\nmid k$, we see that the formula \eqref{thiss} may be rewritten as
\begin{equation}
n_d = \sum_{k|d} \mu(k) \frac{m_{d/k}}{k^s} =
\sum_{\topa{k|d}{p\nmid k}} \frac{\mu(k)}{k^2} \bigl(m_{d/k} - \frac{1}{p^s}m_{d/(pk)}\bigr)
\end{equation}
with the same understanding that $m_{d/(pk)}=0$ if $p\nmid d$.
We see that if \eqref{condition} holds, the $n_d$ are $p$-integral for any $p$, 
hence integral.

We can reformulate the above statement in $p$-adic terms. Let us denote by $V_p(z)$
the $p$-adic reduction of $V(z)$, \ie, the series obtained from $V(z)$ by viewing
all coefficients as $p$-adic numbers. (Really, this is the same series.) We also
denote by ${\rm Fr}_p:\Q[[z]]\to\Q[[z]]$ the Frobenius endomorphism fixing $\Q$ and
sending $z$ to $z^p$. Then the characterization \eqref{fr} of $s$-functions is 
equivalent to the statement that for any $p$,
\begin{equation}
\eqlabel{local}
M_p(z) = \frac{1}{p^s} {\rm Fr}_p V_p(z) - V_p(z)  \in \Z_p[[z]]
\end{equation}
is a series with $p$-adic integral coefficients. We may call a function
that satisfies \eqref{local} (only) for some fixed prime $p$ a local $s$-function.

One can express the coefficients $n_d$ (or, better to say, their 
$p$-adic reductions) in terms coefficients of the series $m_p(z)$, 
see Lemma 3 of \cite{sv2}.)
It follows immediately from this formula that $p$-adic integrality of  coefficients of 
$M_p(z)$ for all $p$ guarantees the integrality of $n_p.$

\section{Regulators.  K-theoretic characterization of 2-functions}

 Fix an odd prime number $p$.
 
Let us define a map 
\begin{equation}\label{eq12}
(f, g)_p:  \Z_p((z))^\times \otimes  \Z_p((z))^\times  \to \Q_p((z))/\Z_p((z))
\end{equation}
to be a unique bilinear skew-symmetric pairing such that for every 
$g\in \Z_p[[z]]^{\times}$ and every $f\in  \Z_p((z))^{\times}$, one has 
\begin{equation}\label{eq5}
 (f, g)_p= \int( \frac{1}{p^2} {\rm Fr}^* (\log g d \log f) - \log g d \log f  ) - 
\frac{1}{p} {\rm Fr}^* (\log g) (\frac 1p {\rm Fr}^* (\log f) - \log f).
\end{equation}
Here  ${\rm Fr}^*$ is the  ``Frobenius lifting'':
$${\rm Fr}^*(h(z))= h(z^p).$$

The key property of the pairing $(f, g)_p$ 
is that it  is invariant under any change of variables of the form $z \to h(z)=a_1 z 
+a_2z^2+\cdots $, $a_1 \in \Z_p^\times$.  In particular, if $(f, g)_p=0$  ({\it i.e.}, 
$(f(z), g(z))_p$ is p-adically integral)  then the same is true  for $(f(h(z)), g(h(z)))_p$.
This can be derived from a more general fact: the right-hand side of formula (\ref{eq5}) 
viewed as an element of the quotient group   $ \Q_p((z))/\Z_p((z))$ does not get changed 
if one replaces ${\rm Fr}^*$ by an arbitrary Frobenius lifting of the form
$${\rm F}^*(h(z))= h(z^p(1+pr(z)),$$
for some $r(z)]\in \Z_p[[z]]$.

One can derive from this fact (or prove directly) that $(1-f, f)_p=0$ for every $f\in  
\Z_p((z))^\times$ such that $1-f$ is also in  $ \Z_p((z))^\times$.
Thus,  (\ref{eq12}) factors through a homomorphism
\begin{equation}\label{eq13}
K^0_2( \Z_p((z))) \to  \Q_p((z))/\Z_p((z)).
\end{equation}
Finally, one can easily check that  $(f, g)_p$  is continuous in each variable with respect 
to ``$z$-adic'' topology on  $  \Z_p((z))$.

The above properties are sufficient to prove Theorem 2.
Indeed, if $Y(z)$ is orthogonal to $z$ then $(z, Y(z))_p=0$ for every odd prime $p$. On 
the other hand, we have that
$$(z, Y(z))_p=  \frac{1}{p^2} {\rm Fr}^* W(z) - W(z).$$
Therefore $W(z)$ is a $2$-function.

\begin{remark} One can check that (\ref{eq13}) factors uniquely through a homomorphism
$$K_2( \Z_p((z))) \to  \Q_p((z))/\Z_p((z)).$$
\end{remark}

\section{Generalizations}
\label{generals}

One of our central contentions in writing this note is that $s$-functions are 
interesting algebraic objects in their own right, independent of relations to 
physics of topological strings. We also claim that the value $s=2$ is special. 
We give further evidence in this section by pointing out some very natural
generalizations of our discussion so far. One possible application that we will 
not (p)review in any detail here is to the theory of Mahler measures. Indeed, the reader 
will find it easy to verify that the relation between the so-called Modular Mahler 
Measures of \cite{frv} and the instanton expansion of certain ``exceptional non-critical 
strings'' \cite{exceptional}, which was pointed out by Stienstra \cite{stienstra}, 
is nothing but an elementary framing transformation \eqref{how1}, \eqref{how2}.
Given our results, it is clear that the relation will be valid in much greater
generality than the examples presented in \cite{stienstra}. It would be 
interesting to explore this further.

\subsection{Arithmetic twists}

The generalization that motivated our initial observations concerning 2-functions is
related to
the results of \cite{arithmetic}. In that work, it was pointed out that the 
A-model instanton expansion of the superpotential associated with a general D-brane 
on a compact Calabi-Yau three-fold is not rational (let alone integral in the
usual sense). Instead, the coefficients were found to be contained in the algebraic
number field $K$ over which the curve representing the D-brane in the B-model
was defined.
However, it was also observed that with an appropriate modification of
the Ooguri-Vafa multi-cover formula, \eqref{total}, at least integrality in the 
algebraic sense could be preserved.
The expansion proposed in \cite{arithmetic} was of the form
\begin{equation}
W(q) = \sum \tilde n_d q^d = \sum n_d {\rm Li}_2^{\rm D} (q^d)
\end{equation}
where $n_d$ are algebraic integers\footnote{Recall that an algebraic number $x$
can be identified with a root of a polynomial $P(x)\in\Q[x]$. If $P$ has coefficients
in $\Z$, leading coefficient $1$, and is irreducible, then $x$ is an algebraic integer.}
and ${\rm Li}_2^{\rm D}$, dubbed the ``D-logarithm'' is a certain (formal) power
series with coefficients in $K$ that depend $n_d$, and
its images under the Galois group of the extension $K/\Q$. More precisely, the
definition of the D-logarithm given in \cite{arithmetic} depended on the lifting 
$\bmod p^2$ of the Frobenius automorphism at each unramified prime $p$, so 
in that sense the coefficients $n_d$ (though integer for any choice of lifting)
depend on several infinities of choices, and would not appear as true geometric 
invariants. The noteworthy exception is provided
by abelian extensions, where ${\rm Li}_2^{\rm D}$ could be taken to be the
di-logarithm twisted by a Dirichlet character $\chi$, of the form
\begin{equation}
\sum \chi(k) \frac{q^k}{k^2}
\end{equation}
As will be clear, this can be rewritten as an integral linear combination of
ordinary di-logarithms evaluated at appropriate roots of unity, which can therefore
be viewed as a canonical basis in which to decompose the superpotential. It remains
rather unclear at this point whether such a basis exists also for general extensions
with non-abelian Galois group.

On the other hand, however,  one may formulate the integrality
statement of \cite{arithmetic} without explicitly referring to any ``D-logarithm''.
Moreover, the proofs of \cite{sv2} can rather straightforwardly be adapted to prove
that integrality statement as well. We will explain this in the forthcoming paper 
\cite{svw2}.

Finally, and this is most relevant given with respect to the present note, 
it turns out that the
framing transformation can also be defined, and preserves integrality (in the
algebraic sense), for 2-functions with coefficients in an arbitrary number field
\cite{svw2}.

\subsection{Multi-variable case}

As we have pointed out before, if $V$ is an $s$-function (for $s>2$), then 
$W_V=\delta^{s-2}V$ is a 2-function, and its framed version $\widetilde {W_V}=
\widetilde{\delta^{s-2}V}$ is also a 2-function. It is a natural question to
ask whether this $2$-function also comes from an $s$-function, namely whether
there exists an $s$-function $\tilde V$ such that $\widetilde{W_V}= \delta^{s-2}
\tilde V$. It is not hard to see that really this is not the case 
(for instance, framing ${\rm Li}_s$ for $s>3$ in this way returns at most a 
3-function). Thus, among $s$-functions for other values of $s$, $2$-functions 
are distinguished by the integrality of framing.

Perhaps the most direct way to see that framing naturally only makes sense for 
$2$-functions is to consider the generalization to the multi-variable situation.
With rational coefficients, we say, as before, 
that a formal power series $W\in \Q[[z_1,\ldots z_r]]$
is a 2-function if it can be written as an integral linear combination of di-logarithms,
\begin{equation}
W(z_1,\ldots, z_r) = \sum_{d_1,\ldots, d_r} n_{d_1,\ldots, d_r} \Li \bigl(z_1^{d_1}
\cdots z_r^{d_r}\bigr)
\end{equation}
Defining $\delta_i\equiv \frac{d}{d\ln z_i}$, and following eq.\ \eqref{introduce}, we introduce
\begin{equation}
\eqlabel{multi}
Y_i = \exp\bigl(\delta_i W\bigr)
\end{equation}
Since the $\delta_i W$ are 1-functions, the $Y_i$ naturally have integer coefficients.
An interesting distinction from the one-variable case is that we find an additional 
degree of freedom when identifying the ``framed'' variables $\tilde z_i$ with $Y_i$, in
analogy to \eqref{how1}. Namely, say $(\kappa^{ij})_{i,j=1,\ldots r}$ is a symmetric 
matrix with integer coefficients. Define $\sigma_i= (-1)^{\kappa_{ii}}$, and
\begin{equation}
\tilde z_i = \sigma_i z_i \prod_{j=1}^r Y_i^{\kappa_{ij}} = 
\sigma_i z_i \exp\bigl(\kappa^{ij}\delta_j W\bigr)
\end{equation}
We may invert this relation as before, and upon writing
\begin{equation}
z_i  = \sigma_i \tilde z_i \exp\bigl(\kappa^{ij}\tilde\delta_j\tilde W\bigr)
\end{equation}
we find that $\tilde W\in\Q[[\tilde z_1,\ldots,\tilde z_r]]$ is also a $2$-function.
This assertion can be proved rather straightforwardly by realizing the multi-dimensional
operation as a combination of elementary one-dimensional framing, leading to the
identification of the group of framing transformations with the additive group of 
symmetric integral matrices.

Now it is clear that if we had started from a multi-variable $s$-function
with $s>2$, we would in \eqref{multi} have obtained more $Y$'s from multi-derivatives than
variables, so the identification would not be one-to-one. Thus, again, $s=2$ is special.

\begin{acknowledgments}
We thank the organizers of String-Math 2012 for the invitation to speak in Bonn,
which led to the present collaboration, and Maxim Kontsevich  for 
valuable discussions and communications.
J.W.\ thanks Fernando Rodriguez-Villegas for drawing attention to ``modular
Mahler measures'' in Summer of 2011, and Henri Darmon for several helpful
conversations. 
The research of J.W.\ is supported in part by an NSERC discovery grant and a
Tier II Canada Research Chair, the research of A. Sch. was supported by NSF grant. 
\end{acknowledgments}





\end{document}